%% file: constraints-final-revise.tex
\documentclass[prd,twocolumn]{revtex4}

\usepackage{graphicx}
\usepackage{color}

\newcommand\ba{{\mathbf{a}}}
\newcommand\bb{{\mathbf{b}}}
\newcommand\be{{\mathbf{e}}}
\newcommand\bff{{\mathbf{f}}}
\newcommand\bx{{\mathbf{x}}}
\newcommand\bxd{\dot{\mathbf{x}}}
\newcommand\bX{{\mathbf{X}}}
\newcommand\bXd{\dot{\mathbf{X}}}
\newcommand\bzero{{\mathbf{0}}}
\newcommand\pa{\partial}
\newcommand\fr{\frac}
\newcommand\beq{\begin{equation}}
\newcommand\eeq{\end{equation}}
\newcommand\bea{\begin{eqnarray}}
\newcommand\eea{\end{eqnarray}}
\newcommand\cd{\cdot}
\newcommand\al{\alpha}
\newcommand\ga{\gamma}
\newcommand\de{\delta}
\newcommand\ep{\epsilon}
\newcommand\tha{\theta}
\newcommand\si{\sigma}
\newcommand\ta{\tau}
\newcommand\ph{\phi}
\newcommand\ps{\psi}
\newcommand\De{\Delta}

\newcommand\cM{{\mathcal M}}
\newcommand{\gsim}{\raise.3ex\hbox{$>$\kern-.75em\lower1ex\hbox{$\sim$}}}
\newcommand{\lsim}{\raise.3ex\hbox{$<$\kern-.75em\lower1ex\hbox{$\sim$}}}

\begin{document}

\title{Constraints on string networks with junctions}

\author{E.~J.~Copeland}
\email{ed.copeland@nottingham.ac.uk} \affiliation{School of
Physics and Astronomy, University of Nottingham, University Park,
Nottingham NG7 2RD, United Kingdom}
\author{T.~W.~B.~Kibble}
\email{kibble@imperial.ac.uk}
\affiliation{Blackett Laboratory, Imperial College, London SW7 2AZ, United Kingdom}
\author{D.~A.~Steer}
\email{steer@apc.univ-paris7.fr} \affiliation{APC, UMR 7164, 10
rue Alice Domon et L\'eonie Duquet, 75205 Paris Cedex 13,  France,
and
\\ LPT, Universit{\'e} de Paris-Sud, Bat.\ 210, 91405  Orsay
Cedex, France.}

\date{\today}

\begin{abstract}

We consider the constraints on string networks with junctions in
which the strings may all be different, as may be found for
example in a network of $(p,q)$ cosmic superstrings. We
concentrate on three aspects of junction dynamics. First we
consider the propagation of small amplitude waves across a static
three-string junction. Then, generalizing our earlier work, we
determine the kinematic constraints on two colliding strings with
different tensions. As before, the important conclusion is that
strings do not always reconnect with a third string; they can pass
straight through one another (or in the case of non-abelian
strings become stuck in an X configuration), the constraint
depending on the angle at which the strings meet, on their
relative velocity, and on the ratios of the string tensions. For
example, if the two colliding strings have equal tensions, then
for ultra-relativistic initial velocities they pass through one
another.  However, if their tensions are sufficiently different
they can reconnect. Finally, we consider the global properties of
junctions and strings in a network. Assuming that, in a network,
the incoming waves at a junction are independently randomly
distributed, we determine the r.m.s.\ velocities of strings and
calculate the average speed at which a junction moves along each
of the three strings from which it is formed. Our findings suggest
that junction dynamics may be such as to preferentially remove the
heavy strings from the network leaving a network of predominantly
light strings.  Furthermore the r.m.s.\ velocity of strings in a
network with junctions is smaller than $1/\sqrt{2}$, the result
for conventional Nambu-Goto strings without junctions in Minkowski
spacetime.
\end{abstract}


\maketitle

\section{Introduction}

The evolution of cosmic string networks and their observational
consequences are attracting a great deal of interest, particularly
since they may lead to indirect observational tests of superstring
theory. The annihilation of two branes at the end of brane
inflation
\cite{Burgess:2001fx,Dvali:2001fw,Jones:2002cv,Kachru:2003aw,Kachru:2003sx}
is thought to lead to the formation of cosmic superstrings which
can be either fundamental F-strings, Dirichlet D-strings, or
$(p,q)$-strings, bound states of the two
\cite{Dvali:2003zj,Copeland:2003bj,Pol,Firouzjahi:2006vp}. Most
importantly, the predicted tensions of these strings are not only
compatible with current observational bounds ($G \mu \lsim
2.3\times 10^{-7}$ using the third year WMAP data
\cite{Seljak:2006bg}), but they also lie in a window which may be
testable with the future LISA gravitational-wave detector
\cite{ViA1,ViA2,Irit}.

Whilst cosmic superstrings share many of the properties of
standard grand unified (GUT) cosmic strings \cite{TomMark,ViSh},
they differ in some important respects. First, when two F-strings
(or two D-strings) intersect, they do not necessarily
`intercommute' --- or exchange partners --- with probability $P=1$
\cite{Jackson:2004zg,Hanany:2005bc}. (An exception for local
cosmic strings is when they intercommute at speeds very close to
the speed of light \cite{Achucarro:2006es}.) The effect of
reducing $P$ is to increase the density of strings in the final
scaling solution and hence the gravity wave signal, though the
exact manner in which this happens is under debate
\cite{ViA1,ViA2,Sakellariadou:2004wq,Avgoustidis:2005nv,Irit}.
Second, the different kinds of strings in a cosmic superstring
network can meet at junctions \cite{Tong}. Thus when an F and
D-sting intersect they cannot intercommute, but rather two
junctions are formed and the original strings become connected by
a third ($p,q$) bound state. Junctions also exist in non-abelian
string networks, for which the fundamental group $\pi_1({\cal M})$
of the vacuum manifold ${\cal M}$ is non-abelian.

Even though several authors have addressed both analytically and
numerically the cosmological evolution of string networks with
junctions
\cite{Vachaspati:1986cc,Spergel:1996ai,Sakellariadou:2004wq,Martins:2004vs,Tye:2005fn,Copeland:2005cy,Saffin:2005cs,Hindmarsh:2006qn},
the late-time behavior of the network is still not fully
understood, particularly when there are different types of string.
One possible outcome is that the junctions play a minor r\^ole and
the network reaches a scaling solution similar to that of a GUT
string network. In that case it is possible to make predictions
for cosmic superstring networks by extracting the information from
the GUT string simulations, with suitable allowance for $P\ne 1$
and in principle for different tensions. A second possibility is
that the presence of junctions makes the network freeze and
eventually dominate the energy density of the universe. Similar
questions have been addressed for networks of domain walls with
junctions
\cite{Bucher:1998mh,Battye:1999eq,Battye:2005hw,PinaAvelino:2006ia,Avelino:2006xy,Carter:2006cf}.

In a previous letter \cite{us} we studied
the dynamics of junctions in a local string network in which the
individual strings have no long-range interactions and are well
described by the Nambu-Goto action. We set up the equations of
motion for three strings of tensions $\mu_j$ ($j=1,2,3$) meeting
at a junction at position $\bX(t)$, and were able to solve for the
dynamics of $\bX(t)$ as well as to determine how the junction
moved along each of the strings. A related approach has also been
developed in the context of representations of baryons as pieces
of open string connected at one common point
\cite{Sharov:1998vy,'tHooft:2004he}. Having constructed this
formalism, we initially presented a simple highly symmetric exact
oscillating loop solution.  As in the case of ordinary cosmic
strings, the existence of exact loop solutions may be important in
analyzing the likely behavior of loops in general. Loops of
strings with junctions generally evolve rather differently to
standard string loops; they do not oscillate periodically and
initial analysis indicates that the number of cusps on the loops
may be rather different than for conventional Nambu-Goto loops. We
intend to return to these points in a future publication.

Our second result concerned the intersection of two strings with
equal tensions $\mu_1$, meeting at an angle $\alpha$, but with
equal and opposite velocities $v$. When these strings collide, for
them to exchange partners and become linked by a third string of
tension $\mu_3$ requires a strong kinematical constraint.  We also
discussed the case of non-abelian strings, which may become joined
by a new string lying along the direction of motion of the two
initial strings, and we found the kinematic constraint for this
case too.

We expect that these kinematical constraints could have significant
consequences on the evolution of string networks with junctions: if the
relative velocity of the strings is large, no junction will form
and the strings will pass through each other.  This in turn means
that fewer loops will be formed, and hence that the network would
radiate less energy.  It may thus be important to include these
constraints in analytic and numerical models of network evolution with
junctions.

In this paper we extend the work of \cite{us}, and focus on three aspects
of junction dynamics which we believe will be important to
understand the global properties of the network. After a review of
the equations of motion describing junction evolution in section
\ref{rev}, we first consider the propagation of small amplitude
waves across a static junction formed of three strings with
generally different tensions (section \ref{ampl}). Such small
waves will inevitably be present on strings in a network, and will
radiate gravitationally. We determine the fraction of energy
transmitted and reflected across the junction. Then, in sections
\ref{inter1} and \ref{inter2}, we generalize the kinematic
constraints of \cite{us} to the case in which the two colliding strings
have different tensions $\mu_1$ and $\mu_2$. Since the initial
velocity of strings is crucial to determining the importance of
the kinematical constraints, in sections \ref{rate} and \ref{asv}
we consider the global properties of junctions and strings in a
network. Assuming that, in a network, the incoming waves at a
junction are randomly distributed, we determine the r.m.s.\
velocities of strings and calculate the average speed at which a
junction moves along each of the three strings from which it is
formed. Our findings suggest that junction dynamics may be such as
to preferentially remove the heavy strings from the network.
Finally we conclude in section \ref{conc}.

\section{Review of the basics}\label{rev}

We begin by briefly reviewing the equations of motion for strings
that form Y junctions obtained in \cite{us}.

We choose the temporal world-sheet coordinate to be the global
time, $\ta=t$, and use the conformal gauge in which the spatial
coordinates $\bx(\si,t)$ satisfy
 \beq \bxd\cd\bx'=0, \qquad \bxd^2+\bx'{}^2=1, \label{gc}\eeq
where $\bxd=\pa_t\bx$ and $\bx'=\pa_\si\bx$.  The action for three
strings of tensions $\mu_j$ ($j=1,2,3$) meeting at a junction is
 \bea S&=&-\sum_j\mu_j\int dt\int d\si\,\tha(s_j(t)-\si)\sqrt{\bx'_j{}^2(1-\bxd_j^2)}
 \nonumber\\
 &&\qquad+\sum_j\int dt\,\bff_j(t)\cd[\bx_j(s_j(t),t)-\bX(t)]
 \label{action} \eea
where $\bX(t)$ is the position of the vertex, $\bff_j$ are
Lagrange multipliers, and the $s_j(t)$ are the values of the
spatial world-sheet coordinate at the vertex.  It is
straightforward to generalize this action to include
other vertices, present for example in loop configurations.

Varying $\bx_j$ yields the wave equation, with solution
 \beq \bx_j(\si,t)=\frac{1}{2}[\ba_j(\si+t)+\bb_j(\si-t)], \eeq
where, in order to satisfy the gauge conditions (\ref{gc}),
 \beq \ba'_j{}^2=\bb'_j{}^2=1. \eeq
The Lagrange multipliers impose boundary conditions that may be written
 \beq \ba_j(s_j+t)+\bb_j(s_j-t)=2\bX(t),\label{abX} \eeq
while varying $\bX$ gives a relation between the Lagrange
multipliers which can be reduced to
 \beq \sum_j\mu_j[(1+\dot s_j)\ba'_j+(1-\dot s_j)\bb'_j]=\bzero.
 \label{sumab}\eeq
Eliminating the $\ba'_j$ using the time derivative of (\ref{abX})
then yields
 \beq \dot\bX=-\fr{1}{\mu}\sum_j\mu_j(1-\dot s_j)\bb'_j, \label{Xb}
 \eeq
where
 \beq \mu= \mu_1+\mu_2+\mu_3.
 \eeq

At the junction, the amplitudes $\bb'_j$ of the incoming waves are
known from the initial conditions; those of the outgoing waves,
$\ba'_j$, together with the position of the vertex, are given by
(\ref{abX}) and (\ref{sumab}), provided the $\dot s_j$ are known.
Finally, the latter may be found by solving the constraint
equations $\ba'_j{}^2=1$.  The result depends on the string
tensions and on the angles between the unit vectors $\bb'_j$.
Relative to \cite{us}, we introduce a slight change of notation, writing
for example $c_1=\bb'_2\cd\bb'_3$ (denoted by $c_{23}$ in \cite{us}).  It
will be useful to introduce the quantities $\nu_j$ defined, for
example, by
 \beq \nu_1=\mu_2+\mu_3-\mu_1. \eeq
Then the value of $\dot s_k$ is given by
 \beq 1-\dot s_k=\frac{\mu M_k(1-c_k)}{\mu_k\cM},
 \label{sdot} \eeq
where the $M_j$ are defined by $M_1=\nu_2\nu_3$ along with two similar equations, and
 \beq \cM=\sum_j M_j(1-c_j). \label{cMdef} \eeq
Notice that since the $\dot s_k \leq 1$, it follows that the string tensions satisfy triangle inequalities,
\beq
\nu_j \geq 0 \label{nuplus}.
\eeq
Finally, it follows from (\ref{sdot}) that the $\dot s_j$ satisfy
the energy conservation condition
 \beq \sum_j \mu_j\dot s_j=0. \eeq

\section{Small-amplitude waves}\label{ampl}

A first interesting application of these results is to the
reflection and transmission of small-amplitude waves at a
junction.  Small amplitude waves are expected to build up on
strings as a result of self-intersections and interactions with
other strings, and they will radiate gravitationally.  It is
important to understand how they propagate across a junction.

Consider three straight, static strings, for which
 \beq \ba'_j=\bb'_j=(\cos\tha_j,\sin\tha_j,0)\equiv\be_j, \eeq
with the junction at $\bX=0$. The equilibrium condition
$\dot{s}_j=0$ determines the angles between them through
(\ref{sdot}): for example, we have
 \bea c_1=\cos(\tha_3-\tha_2)&=&
 \fr{\mu_1^2-\mu_2^2-\mu_3^2}{2\mu_2\mu_3},\nonumber \\
 \sin(\tha_3-\tha_2)&=&
 \fr{\De}{2\mu_2\mu_3}, \eea
where
 \beq \De=\sqrt{\mu\nu_1\nu_2\nu_3}. \eeq
Note that as a consequence of (\ref{nuplus}), $\De$ is real.

Now suppose there is a small incoming perturbation on the first string,
so that $\bb'_1=\be_1+\de\bb'_1$, with
 \beq \de\bb'_1(s)=\ep\bff_1\cos ks,\label{deltab} \eeq
where $\ep$ is a small dimensionless parameter and $\bff_1$ is a unit vector satisfying
$\be_1\cd\bff_1=0$; we set $\de\bb'_2=\de\bb'_3=\bzero$.

The simplest case is where $\bff_1=(0,0,1)$ since then all $\de
c_j=0$, and therefore $\de\dot s_j=0$ so that the junction remains
at the same value of $\sigma$ on all strings.
In that case, the outgoing waves are also in the
$z$-direction. From (\ref{Xb}) we find
 \beq \de\dot\bX(t)=-\fr{\mu_1}{\mu}\ep\bff_1\cos kt\, \eeq
and hence
 \bea
 \de\ba'_1(s)&=&
 \fr{\nu_1}{\mu}\ep\bff_1\cos ks, \nonumber\\
 \de\ba'_2(s)&=&\de\ba'_3(s)=-\fr{2\mu_1}{\mu}\ep\bff_1\cos ks.
 \eea
It is then straightforward to find the fraction of energy
transmitted along the strings of tension $\mu_{2,3}$ and reflected
along the string of tension $\mu_1$.  Since
\beq
\delta E_j = \frac{\mu_j}{4} \int d\sigma (\de \ba'^2_j + \de
\bb'^2_j),
\eeq
we find
\beq
R_1 = \frac{\nu_1^2}{\mu^2}, \qquad T_2 = 4 \frac{\mu_1
\mu_2}{\mu^2} , \qquad T_3 = 4 \frac{\mu_1 \mu_3}{\mu^2}.
\label{RT}
\eeq
For instance, suppose that $\mu_1 \ll \mu_2 \sim \mu_3$, so that
the initial perturbation is along the light string.  Then, as
expected, almost all the energy is essentially reflected off the
junction.

The results are more interesting if we take $\bff_1$ in the
$xy$-plane, because then the values of $s_j$ do oscillate and the
junction no longer stays at a fixed position on each string. Let
us set
 \beq \bff_j=(-\sin\tha_j,\cos\tha_j,0), \eeq
so that $\be_j\cd\bff_j=0$ for each $j$, and again take $\de\bb'_1$
of the form (\ref{deltab}), with $\de\bb'_2=\de\bb'_3=\bzero$.  Then we have
 \bea \de c_1&=&0,\nonumber\\
 \de c_2&=&-\fr{\De}{2\mu_1\mu_3}\ep\cos kt,\\
 \de c_3&=&\fr{\De}{2\mu_1\mu_2}\ep\cos kt.\nonumber \eea
It follows that
 \bea \de\dot s_1&=&\fr{(\mu_2-\mu_3)\nu_1}
 {\De}\ep\cos kt,\nonumber\\
 \de\dot s_2&=&-\fr{\mu_1\nu_1}{\De}\ep\cos kt = - \de\dot s_3.\label{amps-z}
\eea
Despite this difference, the expressions for the outgoing-wave
amplitudes are very similar, except that they are no longer all in
the same direction.  We find
 \beq \de\dot\bX(t)=\ep\cos kt\left(\fr{(\mu_2-\mu_3)\nu_1}{\De}\be_1
 -\fr{\mu_2+\mu_3}{\mu}\bff_1\right) \eeq
from which it follows that
 \bea
 \de\ba'_1(s)&=&
 -\fr{\nu_1}{\mu}\ep\bff_1\cos ks\nonumber\\
 \de\ba'_2(s)&=&\fr{2\mu_1}{\mu}\ep\bff_2\cos ks,\label{amps-xy}\\
 \de\ba'_3(s)&=&\fr{2\mu_1}{\mu}\ep\bff_3\cos ks.\nonumber
 \eea

The fact that the amplitudes of the outgoing waves are the same in
magnitude independent of the orientation of $\bff_1$ is at first
sight remarkable, but actually it follows from the fact that we
can regard the waves as representing massless particles
propagating along the strings.  The amplitudes of the outgoing
waves can be derived from conservation of energy and momentum.  It
follows that the reflection and transmission coefficients are as
given in (\ref{RT}), a result that has also been obtained in the
more general setting of multivolume junctions by
\cite{Fomin:2000tv}.

Comparing (\ref{amps-z}) with (\ref{amps-xy}), one interesting
point emerges: in the first case, $\ba'_1$ is in phase with
$\bb'_1$, whereas in the second case it is in anti-phase.  This
means that if the incoming wave is at an intermediate angle, the
reflected wave is tilted in the opposite direction.

\section{Collisions of straight strings}\label{inter1}

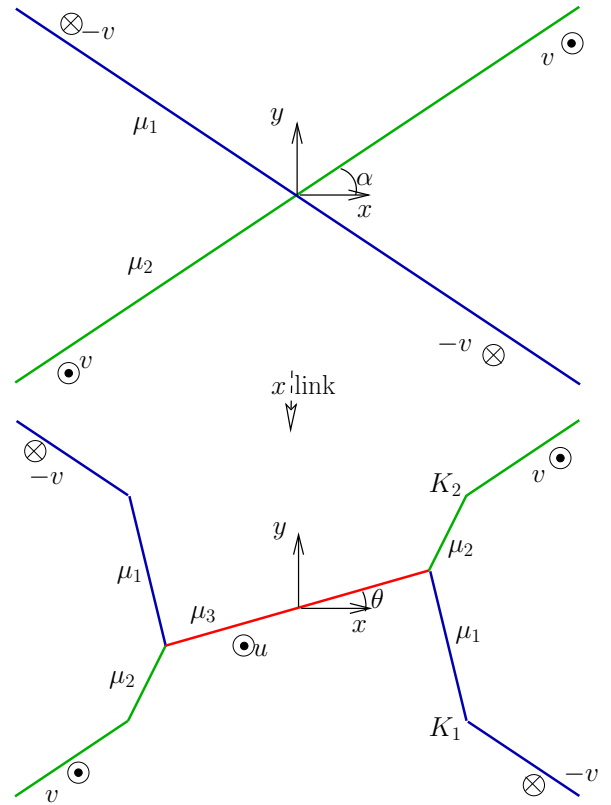
\begin{figure}
\centerline{\scalebox{0.5}{\input{fig1.pstex_t}}} \caption{Two
colliding strings, of unequal tension, joined by a third string
(an $x$-link).} \label{fig:join}
\end{figure}

As summarized in the introduction, in \cite{us} we discussed the
collision of two straight strings of equal tension, $\mu_1=\mu_2$,
and derived kinematical constraints on such a process. Here we
wish to extend this discussion to the case of unequal tension, and
hence it will be useful to define
\beq
\mu_+\equiv\mu_1+\mu_2,\qquad\mu_- \equiv \mu_1 - \mu_2. \label{mumdef}
\eeq
Since $\nu_i \geq 0$, $\mu_3$ is bounded by
\beq
\mu_- \leq \mu_3 \leq \mu_+ \label{superconn}.
\eeq
 This extension is non-trivial as the problem now lacks the
symmetry associated with the equal tension case, and as a first
step one must determine the orientation and velocity of the
joining string after the collision. This is the aim of the present
section.

Consider two strings of tension $\mu_1$ and $\mu_2$ parallel to
the $xy$-plane but at angles $\pm\al$ to the $x$-axis, and
approaching each other with velocities $\pm v$ in the
$z$-direction.  Before the collision (for $t<0$, see Fig.~\ref{fig:join}) we take
 \beq \bx_{1,2}(\si,t)=(-\ga^{-1}\si\cos\al,
 \mp\ga^{-1}\si\sin\al,\pm vt),\label{initial} \eeq
where $\ga^{-1}=\sqrt{1-v^2}$.  Thus
 \bea \ba'_{1,2}&=&(-\ga^{-1}\cos\al,\mp\ga^{-1}\sin\al,\pm v),
 \nonumber\\
 \bb'_{1,2}&=&(-\ga^{-1}\cos\al,\mp\ga^{-1}\sin\al,\mp v). \eea
At the collision, $t=0$, we suppose that the strings interchange
partners. Now, since $\mu_- \neq 0$, we can no longer conclude
that the third -- joining -- string formed after the two original
strings exchange partners, will be static and that it will lie
either along the $x$- or $y$-axis. It is still true that this
third string must be parallel to the $xy$-plane and that there are
two orientations for it, depending on which ends of the original
strings are joined to each other. To be specific, let us assume
that the two ends in the positive-$x$ region are joined at a new
vertex, $\bX$, while those in the negative-$x$ region are joined
to the other end of the new string which has tension $\mu_3$. We
call this an $x$-link --- see Figure \ref{fig:join}.  (The corresponding $y$-link is obtained by
replacing $\alpha \rightarrow (\pi/2 - \alpha)$ and will be
discussed below.)

The new string of tension $\mu_3$ thus lies at an angle $\tha$ to
the $x$ axis, and moves in the $z$-direction with some velocity
$u$; thus
 \beq \bx_3(\si,t)=(\ga_u^{-1}\si\cos\tha,
 \ga_u^{-1}\si\sin\tha,ut),\label{x3} \eeq
where $\ga_u=1/\sqrt{1-u^2}$.  We now determine $\tha$ and $u$. It
follows from (\ref{x3}) that
 \beq \bb'_3=(\ga_u^{-1}\cos\tha,\ga_u^{-1}\sin\tha,-u), \eeq
and thus
 \bea c_1&=&-\ga_u^{-1}\ga^{-1}\cos(\al+\tha)-uv,\nonumber\\
 c_2&=&-\ga_u^{-1}\ga^{-1}\cos(\al-\tha)+uv,\label{cs}\\
 c_3&=&2\ga^{-2}\cos^2\al-1.\nonumber \eea
The position of the vertex is of course $\bX(t)=\bx_3(s_3(t),t)$, so from (\ref{x3}),
 \beq
 \bXd=(\dot s_3\ga_u^{-1}\cos\tha,\dot s_3\ga_u^{-1}\sin\tha,u).
 \eeq
Thus, from (\ref{Xb}) and (\ref{sdot}), we obtain
 \beq \cM\dot\bX+M_3(1-c_3)\bb'_3=
 -M_1(1-c_1)\bb'_1-M_2(1-c_2)\bb'_2, \eeq
whose three components are
\begin{widetext}
 \bea
 [\cM\dot s_3+M_3(1-c_3)]\ga_u^{-1}\cos\tha
 &=&[M_1(1-c_1)+M_2(1-c_2)]\ga^{-1}\cos\al,\nonumber\\
 {[\cM\dot s_3+M_3(1-c_3)]}\ga_u^{-1}\sin\tha
 &=&[M_1(1-c_1)-M_2(1-c_2)]\ga^{-1}\sin\al,\label{simeq}\\
 {[\cM-M_3(1-c_3)]}u
 &=&[M_1(1-c_1)-M_2(1-c_2)]v.\nonumber \eea
\end{widetext}
Dividing the $y$-component of (\ref{simeq}) by the $x$-component,
and comparing with the $z$-component gives
 \beq \fr{\tan\tha}{\tan\al}=\fr{u}{v}=
 \fr{M_1(1-c_1)-M_2(1-c_2)}{M_1(1-c_1)+M_2(1-c_2)}.\label{u/v} \eeq
The first equality here allows us to eliminate $\tha$.  The
second, combined with (\ref{cs}), then allows us to obtain an
equation for $u$, which simplifies to a quadratic for $u^2$ (see
Appendix A):
 \bea \mu_-^2(\sin^2\al) u^4&+&[\mu_3^2(1-v^2)+
 \mu_-^2(v^2\cos^2\al-\sin^2\al)]u^2\nonumber\\
 &-&\mu_-^2v^2\cos^2\al=0. \label{quadratic} \eea
This equation always has one positive root
which then determines
$\tha$ from (\ref{u/v}).  Notice that when $\mu_-=0$ as in \cite{us}, then
$u=0$ and $\theta=0$: this is the symmetric case. On the other
hand, in the limit when the bound in (\ref{superconn}) is saturated, i.e., $\mu_-^2\to\mu_3^2$, then $u\to v$ and $\theta\to\alpha$.  Also, when
$v \rightarrow 1$ for $\mu_- > 0$, then $u \rightarrow 1$ and
again $\theta \rightarrow \alpha$.

The intercommutation produces kinks on the original strings,
moving along them at the speed of light; they are in fact at the
same positions as in the equal tension case, namely at
 \beq K_{1,2}=(\ga^{-1}\cos\al,\pm\ga^{-1}\sin\al,\pm v)t. \eeq

\section{Kinematic constraints}\label{inter2}

Having found the orientation and velocity of the connecting
string, the physical condition that the junctions must move apart
imposes the requirement
\beq \dot s_3>0 \label{cc} \eeq
where $\dot s_3$ is given in (\ref{sdot}). This in turn implies
kinematic constraints on the values of $v$ and $\al$ of the
colliding strings for which an $x$-link can be formed. Replacing
$\al$ by $\pi/2-\al$ gives a similar condition for the formation
of a $y$-link.

When $\mu_-=0$, constraint (\ref{cc}) reproduces the results of \cite{us},
namely
\bea
\al<\arccos\left(\frac{\mu_3\ga}{2\mu_1}\right) &&
(x\mathrm{-link}) \label{k1} \\
\al>\arcsin\left(\frac{\mu_3\ga}{2\mu_1}\right)&&
(y\mathrm{-link}). \label{k2}
\eea
If these bounds are not satisfied, then abelian strings presumably
pass through one another.  Essentially this same result, in the
special case of all tensions equal $\mu_1=\mu_2=\mu_3$, was
obtained earlier in the context of hadronic strings \cite{Artru},
and also when $\mu_-=0$ from a different starting point in the
context of Type-I strings \cite{BK}.

For non-abelian strings, as discussed in \cite{us}, there is a third
possibility namely that a link forms in the direction of motion of
the two initial strings (a $z$-link). Indeed if the string fluxes
on strings 1 and 2 do not commute, then for topological reasons
the strings cannot simply pass through one another: they may only
do so with the formation of a third string. Note that in this case
we do not have strings with three different tensions meeting at a
vertex: at one end of string 3 it is attached to two segments of
string 1, and at the other to two segments of string 2.  Hence the
linking string in this case \emph{does} lie along the $z$-axis,
simplifying the analysis. In effect, we have two vertices of the
type discussed in \cite{us}. The triangle inequalities (\ref{nuplus}) now
require that both ${\mu_3} \leq 2 \mu_1$ and $\mu_3 \leq {2
\mu_2}$. Furthermore the kinematic constraint in this case is that
the {\it total} length of string 3 must increase; we must add the
rates from the two ends. It is not absolutely necessary that $\dot
s_3$ at one particular vertex should be positive.  The required
condition is
 \beq \fr{2\mu_1v-\mu_3}{2\mu_1-\mu_3v}+
 \fr{2\mu_2v-\mu_3}{2\mu_2-\mu_3v}>0. \eeq
Solving for $v$ gives the limit
 \beq v>\fr{4\mu_1\mu_2+\mu_3^2-\sqrt{(4\mu_1^2-\mu_3^2)
 (4\mu_2^2-\mu_3^2)}}{2(\mu_1+\mu_2)\mu_3}. \label{zcon}\eeq

\begin{figure}
\includegraphics[width=0.47\textwidth]{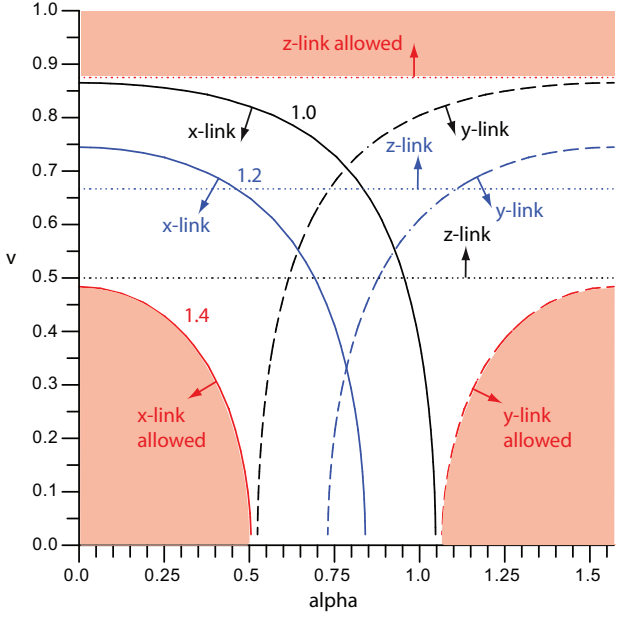}
\caption{Kinematic constraints for $\mu_-=0$.  Allowed regions for $x$-links are to the left of the full curves; for $y$-links to the right of the dashed curves; and, for $z$-links in the non-abelian case, above the horizontal dotted lines.  The values of $\mu_3$ are 1.4 (red), 1.2 (blue), 1.0 (black).  Allowed regions are shaded for the $\mu_3=1.4$ case.}
\label{fig:eqmu}
\end{figure}

\begin{figure}
\includegraphics[width=0.47\textwidth]{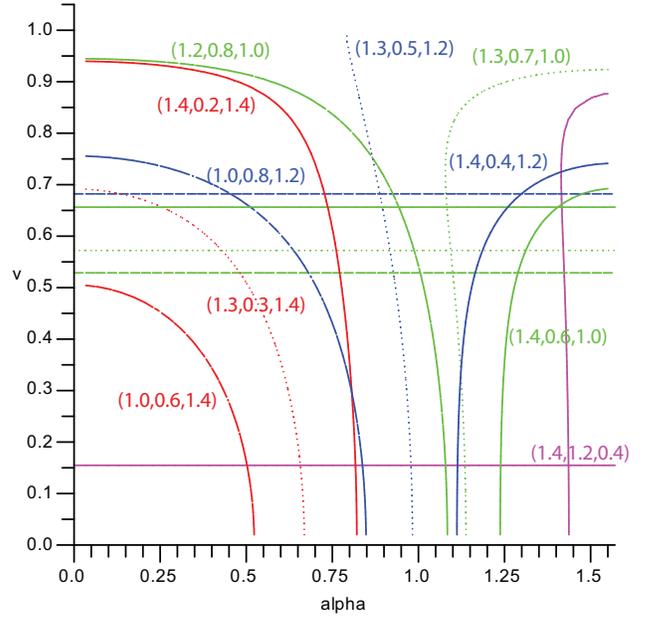}
\caption{Kinematic constraints for strings of unequal tensions
normalized to $\mu=3$ for the formation of $x$- and $z$
links. The allowed region is to the left of the $x$-link curves.  For $z$-links, it is above the horizontal line (which is absent if $\mu_3>2\mu_2$).  The values of $\mu_3$ are again color-coded.  The values of $(\mu_1,\mu_2,\mu_3)$ shown are: $(1.0,0.6,1.4)$, red, dashed; $(1.3,0.3,1.4)$, red, dotted; $(1.4,0.2,1.4)$, red, solid; $(1.0,0.8,1.2)$, blue, dashed; $(1.3,0.5,1.2)$, blue, dotted; $(1.4,0.4,1.2)$, blue, solid; $(1.2,0.8,1.0)$, green, dashed; $(1.3,0.7,1.0)$, green, dotted; $(1.4,0.6,1.0)$, green, solid; $(1.4,1.2,0.4)$, magenta, solid.}
\label{fig:uneqmu}
\end{figure}

These different constraints are summarized in Figures
\ref{fig:eqmu} and \ref{fig:uneqmu} showing the $(\alpha,v)$ plane
for $\mu=3$.  Fig.~\ref{fig:eqmu} shows the three constraints for
the case $\mu_-=0$ (equations (\ref{k1}), (\ref{k2}) and
(\ref{zcon})) and various values of $\mu_3$ (distinguished by
color). For non-abelian strings there are three different regimes:
{\it i}) if $\mu_3 > \sqrt{2}\mu_1$ there is a kinematically
forbidden region for all $\alpha$; {\it ii}) for $ 2\mu_1/\sqrt{3}
> \mu_3 > \sqrt{2}\mu_1$ only a restricted range of $\alpha$ are
forbidden; {\it iii}) for $ \mu_3 < 2 \mu_1/\sqrt{3}$ the whole
$(\alpha,v)$ plane is included in at least one of the allowed
regions.  Allowed regions for the formation of an $x$-link are to
the left of the solid lines, those for a $y$-link to the right of
the dashed lines, while for a $z$-link in the non-abelian case
they are above the horizontal dotted lines.

Fig.~\ref{fig:uneqmu} shows the allowed regions for a number of
cases with $\mu_->0$.  (The sign of $\mu_-$ does not affect the
limits.)  The allowed $x$-link regions are to the left of the
curves, and those for a $z$-link above the horizontal lines.  (The
$y$-link regions can be found by the substitution
$\al\to\pi/2-\al$.)

When $\mu_- \neq 0$, it does not seem to be possible to solve
(\ref{cc}) analytically for all $\alpha$.
We note, however, that there are certain limiting cases for which
an analytic solution is possible.  Firstly, in the low-velocity
limit, $v\to0$, it is clear from (\ref{quadratic}) that also
$u\to0$.  In this limit, $\sin\tha=(\mu_-/\mu_3)\sin\al$.  It is
then straightforward to show that an $x$-link is only possible if
 \beq
 \sin^2\al_{(v=0)}<\fr{\mu_+^2-\mu_3^2}{\mu_+^2-\mu_-^2}.
 \eeq
Note that the value of $\al$ increases if $\mu_3$ decreases, or if $\mu_-$ increases.  [Recall that the triangle inequalities impose the restrictions (\ref{superconn}).]

We can also find a solution
for $\alpha=0$.  Here $u$ is readily obtainable from
(\ref{quadratic}) and then (\ref{cc}) yields
\beq
v_{(\alpha=0)}^2 <
\frac{\mu_3^2(\mu_+^2-\mu_3^2)}{\mu_+^2(\mu_3^2-\mu_-^2)} \equiv
v_c^2. \label{vc}
\eeq
Thus $v_c$ defined in (\ref{vc}) increases with $\mu_-^2$.  Both these limits indicate that in general the kinematic constraints exclude a
smaller region of the $(\alpha,v)$ plane as the string tensions
become more different.  This behavior is readily seen in Fig.~\ref{fig:uneqmu}.

In the equal tension case, $\mu_-=0$, $v_c$ is always less than 1.  Here, however, $v_c^2 \leq 1$ only if
\beq
\mu_3^2 \geq \mu_+ \mu_- = |\mu_1^2 - \mu_2^2|. \label{rel}
\eeq
If this condition is satisfied by the tension of the potentially
linking string, there is a velocity $v_c$ above which abelian
strings will necessarily pass through each other rather than
intercommuting.  However, when the condition is violated, the
entire high-velocity region $v\to1$ for all $\al$ is included in
the allowed region.  The bounding curve in this case bends to the
right, and reaches $\al=\pi/2$ at a finite velocity, above which
an $x$-link is kinematically allowed for any angle $\al$.  This
velocity constraint for $\al=\pi/2$ is
\beq
v_{(\al=\pi/2)}^2 >
\frac{\mu_+^2(\mu_3^2-\mu_-^2)}{\mu_3^2(\mu_+^2-\mu_3^2)}.
\label{vc2}
\eeq
Note that this limit is equal to $1/v_c^2$.

\section{Rate of change of string lengths}\label{rate}

One of the important reasons for studying the kinematics of string
collisions is that the results may throw some light on the
question of how a network of such strings would evolve in the
early universe.  If we ignore the Hubble expansion and any energy
loss mechanisms, then the energy in the string network is fixed,
but some strings will shorten and others will grow.  We may ask how fast,
on average, is this growth or shortening.

It is reasonable to assume that at any string junction, the unit
vectors $\bb'_j$ representing the ingoing waves are randomly
distributed on the unit sphere, and mutually independent.  (This
might not be true if for example two of the strings come from the
same other vertex, but that is presumably not a common situation.)
If the strings are all of the same tension, then because of energy
conservation it is clear that $\langle\dot s_j\rangle$ must vanish
for each $j$, but this is not necessarily so if the tensions are
different.  And as we shall see, even for the equal-tension case,
the zero mean does not mean that the distribution is symmetrical;
it is actually not true that strings are as likely to grow as to
shrink.

Let us start by looking at the distribution of the variables
$c_j$, assuming that the unit vectors $\bb'_j$ are randomly
distributed on the unit sphere, and mutually independent. Let us
choose the $z$-axis along the direction of $\bb'_3$, and $\bb'_2$
in the ($x,z)$ plane; $\bb'_2=(\sin \theta,0,\cos \theta)$.  Then
$\bb'_1=(\sin\beta \cos \phi, \sin \beta \sin \phi,\cos
\beta)$ and we may assume a uniform distribution in the
variables $c_1,c_2$ and $\ph$ where
 \beq c_3=c_1c_2+\sqrt{1-c_1^2}\sqrt{1-c_2^2}\cos\ph.\label{c3phi}
  \eeq
Our aim is to calculate the probability distribution for, say, the rate
$\dot s_1$ at which the first string grows.
Specifically, let
$P(\dot s_1)d\dot s_1$ be the probability that $\dot s_1$ lies between
$\dot s_1$ and $\dot s_1+d\dot s_1$.  Clearly,
\bea P(\dot s_1)&=&\fr{1}{8\pi}\int_{-1}^{1}dc_1\int_{-1}^{1}dc_2
 \int_0^{2\pi}d\ph \nonumber \\
 &&\times\de\left(\dot s_1-1+\fr{\mu M_1(1-c_1)}{\mu_1 \cM}\right),
 \label{Pw1}
 \eea
where we have used (\ref{sdot}), and $\cM=\cM(c_1,c_2,c_3)$ is
defined in (\ref{cMdef}).

\begin{figure}
\includegraphics[width=0.47\textwidth]{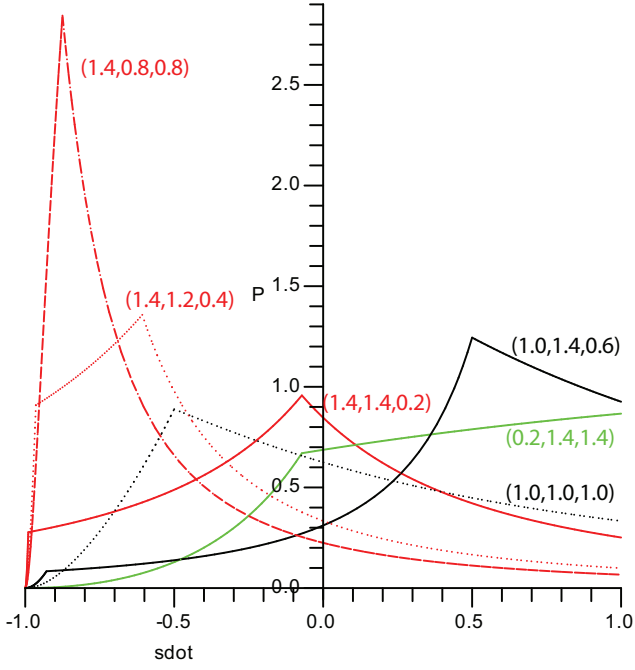}
\caption{The distribution of $P(\dot{s}_1$) plotted against $\dot{s}_1$.  The values of $\mu_1$ are indicated by colors.  The curves shown are for $(\mu_1,\mu_2,\mu_3)=(1.4,0.8,0.8)$, red, dashed; $(1.4,1.2,0.4)$, red, dotted; $(1.4,1.4,0.2)$, red, solid; $(1.0,1.0,1.0)$, black, dotted; $(1.0,1.4,0.6)$, black, solid; $(0.2,1.4,1.4)$, green, solid.}
\label{fig:sdotdist}
\end{figure}

The resulting calculation is fairly lengthy, so we relegate the details to Appendix B.  The distribution $P(\dot s_1)$ takes three different analytic forms in three intervals, namely
 \bea
 1.&& -1<\dot s_1<\fr{\mu_3^2-\mu_1^2-\mu_2^2}{2\mu_1\mu_2};
 \nonumber\\
 2.&&\fr{\mu_3^2-\mu_1^2-\mu_2^2}{2\mu_1\mu_2}<\dot s_1<
 \fr{\mu_2^2-\mu_1^2-\mu_3^2}{2\mu_1\mu_3}; \label{regime}\\
 3.&&\fr{\mu_2^2-\mu_1^2-\mu_3^2}{2\mu_1\mu_3}<\dot s_1<1.
 \nonumber
 \eea
(Without loss of generality, we have chosen $\mu_2\ge \mu_3$.)
To write the final expressions for the probability distribution
concisely, it is useful to introduce the constant
 \beq Q=\fr{\mu}{3\nu_1\nu_2\nu_3}. \eeq
Then the expressions for $P(\dot s_1)$ in the three regions are
\begin{widetext}
 \bea
 1.&&  P(\dot s_1)=Q\mu^2\mu_1^2
 \left(\fr{1+\dot s_1}{1-\dot s_1}\right)^2
 \fr{(2\mu_2+2\mu_3-\mu_1)+(2\mu_1-\mu_2-\mu_3)\dot s_1}
 {(\mu_1\dot s_1+\mu_2+\mu_3)^3};
 \nonumber\\
 2.&& P(\dot s_1)=Q\nu_2^2\fr{(2\mu_1+\mu_2-\mu_3)}
 {\mu_1(1-\dot s_1)^2};\label{ranges}\\
 3.&& P(\dot s_1)=Q\mu_1^2\nu_1^2
 \fr{(\mu_1+2\mu_2+2\mu_3)+(2\mu_1+\mu_2+\mu_3)\dot s_1}
 {(\mu_1\dot s_1+\mu_2+\mu_3)^3}.\nonumber
 \eea
\end{widetext}
The probability distribution has kinks at the boundaries between
these regions, given by (\ref{ranges}). The form of the
distribution is illustrated for various cases in
Fig.~\ref{fig:sdotdist}.

In the particular case where the tensions are all equal, the
distribution has a single kink, at $\dot s_1=-\fr{1}{2}$.  Its
form is
 \bea
 1.&&  P(\dot s_1)=\fr{27}{(\dot s_1+2)^3}
 \left(\fr{1+\dot s_1}{1-\dot s_1}\right)^2
 \qquad\left(\dot s_1<-\fr{1}{2}\right) \nonumber\\
 3.&& P(\dot s_1)=\fr{4\dot s_1+5}{(\dot s_1+2)^3}
 \qquad\left(\dot s_1>-\fr{1}{2}\right).\nonumber
 \eea
It is interesting that although the mean value of $\dot s_1$ in
this distribution is zero, as it must be, the distribution is not
symmetrical --- the dotted black curve in Figure \ref{fig:sdotdist}.  At any particular time, it is most probable that one of the three legs is growing, while the other two are shrinking (of course at a slower rate).

More generally, if $\mu_2=\mu_3$, there is only a single kink; three of the curves in Fig.~(\ref{fig:sdotdist}) are examples of this case, with $\mu_1=1.4$, 1.0, and 0.2.

It is now straightforward (if tedious) to compute the average
value of $\dot s_1$.  We find
 \bea
 \langle\dot s_1\rangle&=&\fr{3\mu_1-\mu}{3\mu_1}
 +\fr{Q}{\mu_1}\biggl[
 -(\mu_1+\mu)\nu_1^2
 \ln\fr{\mu\nu_1}{4\mu_2\mu_3}\nonumber\\
 &&+(\mu_1+\nu_3)\nu_2^2
 \ln\fr{\mu\nu_2}{4\mu_1\mu_3}\nonumber\\
 &&+(\mu_1+\nu_2)\nu_3^2
 \ln\fr{\mu\nu_3}{4\mu_1\mu_2}\biggr].\label{Psdot}\\ \nonumber
 \eea
Note that the expression (\ref{Psdot}) is symmetrical under the
interchange $\mu_2\leftrightarrow\mu_3$, as it should be.  It is
also easy to check that it satisfies the consistency condition
 \beq \mu_1\langle\dot s_1\rangle+\mu_2\langle\dot s_2\rangle
 +\mu_3\langle\dot s_3\rangle=0. \eeq

\begin{figure}
\includegraphics[width=0.47\textwidth]{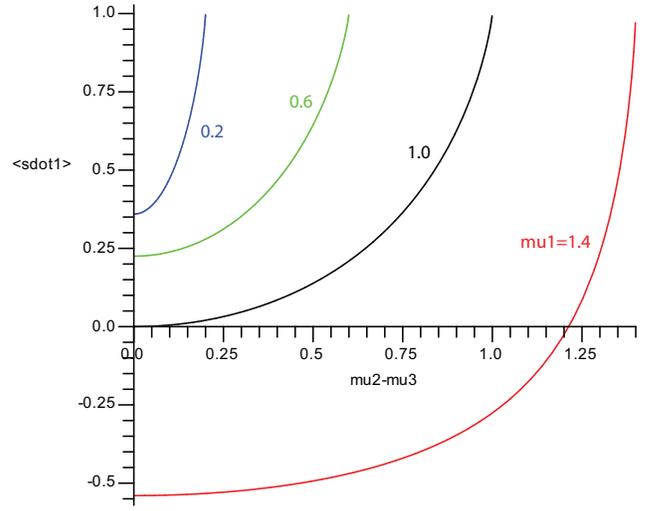}
\caption{Average value of $\dot s_1$ plotted against $|\mu_2-\mu_3|$
for $\mu_1=0.2$ (blue), 0.6 (green), 1.0 (black) and 1.4 (red).} \label{fig:sdotav}
\end{figure}

Finally it is interesting to note that in general $\langle\dot
s_1\rangle$ is more likely to be positive if $\mu_1$ is small, or
if the other two tensions are very different. In
Fig.~\ref{fig:sdotav}, the average value is plotted against
$|\mu_2-\mu_3|$ for various values of $\mu_1$.  Notice that for
$\mu_1\le 1$, $\langle\dot s_1\rangle$ is \emph{always} positive.
It appears that in a network of strings there may be tendency for
the lighter strings to grow at the expense of the heavier ones.
This seems to be consistent with results obtained by studying the
statistical mechanics of such networks of strings \cite{RS}.
\vskip 1cm

\section{Average string velocity}\label{asv}

For ordinary Nambu-Goto strings in flat space-time, the r.m.s.\
transverse string velocity in a random tangle of strings is
$1/\sqrt2$.  It is interesting to ask whether this figure would be
different for junction-forming strings.  It is easy in principle
to answer this question.  For example the square of the velocity
of the first string adjacent to the vertex is
 \beq \dot\bx_1^2=\fr{1}{2}(1-\ba'_1\cd\bb'_1), \eeq
and using (\ref{abX}) and (\ref{Xb}) this can be expressed in
terms of the $c_j$.  We have
 \bea (1+\dot s_1)\ba'_1\cd\bb'_1&=&\fr{\nu_1}{\mu}(1-\dot s_1)
 -\fr{2\mu_2}{\mu}(1-\dot s_2)c_3\nonumber\\
 &&-\fr{2\mu_3}{\mu}(1-\dot s_3)c_2.\eea
Using (\ref{sdot}) this becomes
\begin{widetext}
 \beq \ba'_1\cd\bb'_1=\fr{\nu_1M_1(1-c_1)-2\mu_1M_2(1-c_2)c_3
 -2\mu_1M_3(1-c_3)c_2}
 {-\nu_1M_1(1-c_1)+2\mu_1M_2(1-c_2)+2\mu_1M_3(1-c_3)}. \eeq
\end{widetext}

We can then compute the average over the probability distribution
already obtained by plugging this expression into an integral of
the same form as (\ref{Pw1}).  In this case, there does not seem
to be any obvious way of obtaining an analytic result, but it is
possible to make some progress.  It seems to be slightly easier to
compute the average of $\bx'{}^2$ rather than $\dot\bx^2$:
 \bea \langle \bx'_1{}^2 \rangle &=&
 \fr{1}{2} \langle1+\ba'_1\cd\bb'_1 \rangle \nonumber\\
 &=&\left\langle\fr{2\mu_1^2(1-c_2)(1-c_3)}
 {-M_1(1-c_1)+2\mu_1[\nu_3(1-c_2)+\nu_2(1-c_3)]}\right\rangle.\nonumber\\ \eea

Then if we use as independent variables $c_2,c_3,\ps$ with now
 \beq c_1=c_2c_3+\sqrt{1-c_2^2}\sqrt{1-c_3^2}\cos\ps,\label{c1psi}
  \eeq
we can perform the $\ps$ integration, to yield
 \beq \langle\bx'_1{}^2\rangle=\fr{\mu_1^2}{2}\int_{-1}^1 dc_2
 \int_{-1}^1 dc_3\fr{(1-c_2)(1-c_3)} {\sqrt{F^2-G^2}}, \label{kyoto} \eeq
where
 \beq F=2\mu_1[\nu_3(1-c_2)+\nu_2(1-c_3)]-\nu_2\nu_3(1-c_2c_3), \eeq
and
 \beq G=\nu_2\nu_3\sqrt{1-c_2^2}\sqrt{1-c_3^2}. \eeq
Remarkably (\ref{kyoto}) can be evaluated to give the general
result
\bea \langle\dot{\bf x}_1^2\rangle &=&  \langle (1- \bx'_1{}^2)\rangle \nonumber \\
&=&
 \frac{\mu_1^2-13\bar{\mu}^2}{15(\mu_1^2-\bar{\mu}^2)}+
 \frac{(4\mu_1-\bar{\mu})(\mu_1+\bar{\mu})^2}{15\mu_1(\mu_1-\bar{\mu})^2}
 \ln\frac{2\mu_1}{\mu_1+\bar{\mu}} \nonumber \\
 &~&
 + \frac{(4\mu_1+\bar{\mu})(\mu_1-\bar{\mu})^2}{15\mu_1(\mu_1+\bar{\mu})^2}
 \ln\frac{2\mu_1}{\mu_1-\bar{\mu}}, \label{xdotsq}\eea
where we have introduced $\bar{\mu}=\mu_2-\mu_3$. A number of
interesting features can be seen.  First, the result is
independent of $\mu_2+\mu_3$, depending only on the single ratio
$\bar{\mu}/\mu_1$.  Secondly, the limit $\bar{\mu}=0$ gives
 \beq
 \langle\dot{\bf x}_1^2\rangle = \frac{1}{15} \left[1 + 8\ln2 \right] \simeq
 0.436,
 \eeq
 independently of $\mu_1$. In particular we note that, even
for strings of equal tension, the r.m.s.\ velocity is not
$1/\sqrt2$ as it is for ordinary Nambu-Goto strings without
junctions.  Surprisingly, on the contrary, this value is obtained
when $\bar{\mu}=\mu_1$, a limit in which the triangle inequality
is only just satisfied.

\begin{figure}
\includegraphics[width=0.47\textwidth]{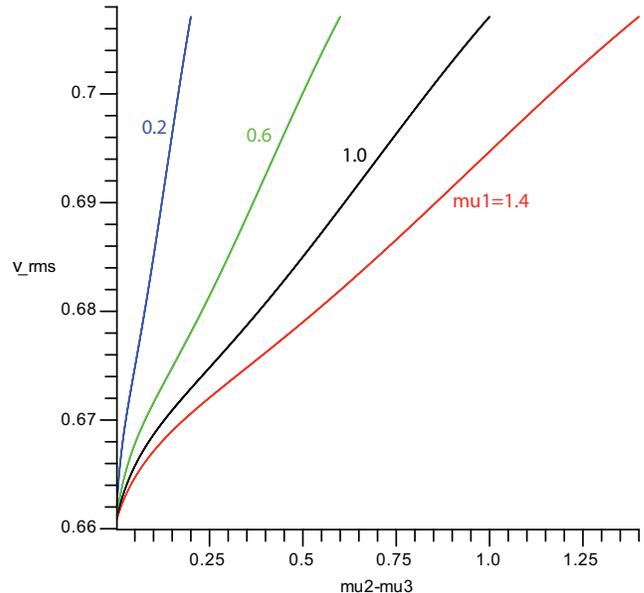}
\caption{R.m.s.\ value of the string velocity $v$ plotted
against $|\mu_2-\mu_3|$ for several values of $\mu_1$.} \label{fig:vav}
\end{figure}

For completeness, in Fig.~\ref{fig:vav} we have plotted $v_{\rm
rms} = \sqrt{\langle\dot{\bf x}_1^2\rangle}$, for various choices
of $\mu_1$, a function of $|\mu_2-\mu_3|$.

\section{Conclusions}\label{conc}

In an earlier letter \cite{us}, we considered  the kinematic
constraints on the possibility of intercommuting of strings that
form junctions. Concentrating on the case where the approaching
strings have equal tensions, we showed that if the relative
velocity with which they meet is too large, no exchange can take
place and the strings will pass through one another (or, for
non-abelian strings, become joined by a string in the direction of
the relative velocity or form a linked X configuration).  In this
paper we have extended the analysis to the more realistic case
where the approaching strings have different tensions.  For some values of the string tensions, it may happen that ultra-relativistic strings \emph{can} exchange partners.

We first studied the reflection and transmission of small-amplitude waves at string junctions --- something that may be important in predicting the gravitational radiation from strings.  We
determined the fraction of energy transmitted and reflected across
the junction, and showed how the reflected wave is generally
tilted in the opposite direction to the incoming wave.

In sections \ref{inter1} and \ref{inter2}, we generalized the
kinematic constraints (\ref{k1}) and (\ref{k2}) of I to the case
in which the two colliding strings have different tensions $\mu_1$
and $\mu_2$. The question of whether strings always intercommute
is vital in understanding the evolution of a network of strings;
in particular it affects the final density of strings found in the
scaling regime of a network, if indeed a scaling solution is
reached.  We have established the criteria required for
intercommuting in terms of the incoming velocity and angle of
approach, as a function of the string tensions.  A particularly
important combination of tensions is that given in the inequality
(\ref{rel}).  If this inequality is satisfied, for instance when
$\mu_1=\mu_2$, then ultra-relativistic strings cannot
intercommute. If it is violated, then ultra-relativistic strings
may intercommute.  Our result that intercommuting does not always
happen appears at first sight to be in contrast to the claim in
\cite{Eto:2006db}, but we believe the regime where their results
are applicable corresponds to low velocities of approach as the
moduli approximation they use breaks down for high velocities. It
is in the high velocity regime that intercommutation may break
down.

Since the initial velocity of colliding strings is crucial to
determining the importance of the kinematical constraints, in
sections \ref{rate} and \ref{asv} we considered the global
properties of junctions and strings in a network. A plausible
assumption for a string network is that the incoming waves at a
junction are randomly distributed. This has allowed us to make
progress in determining the r.m.s.\ velocites of strings. In
particular (ignoring energy loss mechanisms such as expansion of
the universe) we have calculated the average speed at which a
junction moves along each of the three strings from which it is
formed. Our results are intriguing. For example, even for the case
of equal tension strings, although the average velocity of the
junctions is zero as expected, the distribution of the velocities
is not peaked around zero, but around a negative velocity
(actually around $\dot{s_i} = -1/2$) indicating that even in this
apparently symmetric case, it is most probable that at any
particular time, one of the three legs is growing, while the other
two are shrinking, all be it at a slower rate. The case of unequal
tensions can also be solved analytically and our results suggest
that junction dynamics may be such as to preferentially remove the
heavy strings from the network. It thus seems likely that the
system will evolve to one where the lightest strings are
dominating the dynamics, though of course junctions are still
present. Regarding the r.m.s.\ velocities of the strings
themselves, we showed that they are generically smaller than the
$1/\sqrt{2}$ characteristic of Nambu-Goto networks, even in the
case when the strings all have equal tensions.

In a future publication we intend to present some exact solutions
for loops containing junctions. Given the new features we have
uncovered for the dynamics of string networks with junctions, we
should not be surprised to find important results concerning the
distribution of kinks and cusps in these more complicated
configurations. This in turn could have a bearing on the
gravitational radiation emitted from loops of strings with
junctions.

\begin{acknowledgments}
The work reported here was assisted by the ESF COSLAB Programme.
DAS is grateful to the Physics Department at Case Western for
hospitality whilst some of this work was done.  We are grateful to
Kepa Sousa for a useful comment, and also Xavier Artru for drawing
reference \cite{Artru} to our attention.
\end{acknowledgments}

\section*{APPENDIX A}

Here we outline the calculation of equation (\ref{quadratic}).
Let $\mu_\pm=\mu_1\pm\mu_2$, so that
 $$M_{1,2}=(\mu_+-\mu_3)(\mu_3\pm\mu_-),\qquad M_3=\mu_3^2-\mu_-^2.$$

Then the second equality in (\ref{u/v}) yields
 \bea \frac{u}{v}= \qquad \qquad \qquad \qquad \qquad \qquad
 \qquad \qquad \qquad \qquad \qquad
 \nonumber \\ \frac{\mu_-(1+\ga^{-1}\ga_u^{-1}\cos\al\cos\tha)
 +\mu_3(uv-\ga^{-1}\ga_u^{-1}\sin\al\sin\tha)}
 {\mu_3(1+\ga^{-1}\ga_u^{-1}\cos\al\cos\tha)
 +\mu_-(uv-\ga^{-1}\ga_u^{-1}\sin\al\sin\tha)}.
\nonumber
  \eea Next we can eliminate $\cos\tha$ and $\sin\tha$.
>From the first equality in (\ref{u/v}),
 \bea
 \cos\tha&=&\frac{v\cos\al}{\sqrt{v^2\cos^2\al+u^2\sin^2\al}},\nonumber
 \\
 \sin\tha&=&\frac{u\sin\al}{\sqrt{v^2\cos^2\al+u^2\sin^2\al}}. \nonumber
 \eea
Thus multiplying up and grouping all the terms involving
$\ga^{-1}$ on the right, gives
 \bea && (\mu_3u-\mu_-v)+(\mu_-u-\mu_3v)uv=
 \nonumber \\
&&\frac{\ga^{-1}\ga_u^{-1}}{\sqrt{v^2\cos^2\al+u^2\sin^2\al}}
[(\mu_-v-\mu_3u)v\cos^2\al+ \nonumber \\
&&  \qquad \qquad \qquad (\mu_-u-\mu_3v)u\sin^2\al]. \nonumber
\eea Now square to obtain
 \bea &&(v^2\cos^2\al+u^2\sin^2\al)[\mu_3u(1-v^2)-\mu_-v(1-u^2)]^2 =  \nonumber\\
 &&-(1-v^2)(1-u^2)[\mu_-(v^2\cos^2\al+u^2\sin^2\al)-\mu_3uv]^2. \nonumber \eea
On expansion, the terms in $uv$, which come from the cross terms in
each square bracket, cancel, and we are left with
$$
F\left[
\mu_3^2u^2(1-v^2)-\mu_-^2(1-u^2)(v^2\cos^2\al+u^2\sin^2\al)
\right] =0
$$
 where
$$
F = (1-v^2)(v^2\cos^2\al+u^2\sin^2\al)-v^2(1-u^2) \nonumber
$$
which vanishes only if $v=\pm u$.  Thus we finally get the
quadratic (\ref{quadratic}) for $u^2$:
 \bea \mu_-^2(\sin^2\al) u^4&+&[\mu_3^2(1-v^2)+
 \mu_-^2(v^2\cos^2\al-\sin^2\al)]u^2\nonumber\\
 &-&\mu_-^2v^2\cos^2\al=0. \nonumber \eea
 This always has one
positive root as the discriminant is positive.

\section*{APPENDIX B}

To carry out the integral in Eq.~(\ref{Pw1}),
it is convenient to go over to a more symmetrical form by
changing variable from $\ph$ to $c_3$, using (\ref{c3phi}).  This
introduces a Jacobian factor, which is the inverse of
 $$\fr{dc_3}{d\ph}=-\sqrt{1-c_1^2}\sqrt{1-c_2^2}\sin\ph
 =-\sqrt J, $$
where
 \bea J&=&(1-c_1^2)(1-c_2^2)-(c_3-c_1c_2)^2\nonumber\\
 &=&1-c_1^2-c_2^2-c_3^2+2c_1c_2c_3. \nonumber\eea
Note that the physically allowed region in $(c_1,c_2,c_3)$ space
is characterized by $J\ge0$.  There is also a factor of $2$
because there are two values of $\ph$ for each $c_3$.  Letting $w_1=1-\dot s_1$, we arrive at
 \bea P(1-w_1)&=&\fr{1}{4\pi}\int_{-1}^{1}dc_1\int_{-1}^{1}dc_2
 \int_{-1}^{1}dc_3\fr{\tha(J)}{\sqrt{J}}\nonumber\\
 &&\times\de\left(w_1-\fr{\mu M_1(1-c_1)}{\mu_1 \cM}\right). \eea

It is straightforward to perform the $c_3$ integral using the
delta function.  This gives
 \beq P(1-w_1)=\fr{1}{4\pi} \fr{\mu}{\mu_1} \fr{M_1}{M_3} \fr{1}{w_1^2}
 \int_{-1}^{1}dc_1 (1-c_1) \int_{-1}^{1}dc_2
 \fr{\tha(J)}{\sqrt{J}}, \label{Pw1a}\eeq
where now
 \beq J=(1-c_1^2)(1-c_2^2)-[c_3(w_1)-c_1c_2]^2.\label{J} \eeq
Here $c_3(w_1)$ is obtained by equating to zero the argument of
the delta function in (\ref{Pw1a}) and solving for $c_3$:
 \beq
 c_3(w_1) = 1 + \fr{M_1}{M_3}(1-c_1)
 \left(1-\fr{\mu}{\mu_1 w_1}\right) + \fr{M_2}{M_3}(1-c_2).
 \eeq
Now since, by (\ref{J}), $J$ is a quadratic function of $c_2$, we
can go on to perform the $c_2$ integral in (\ref{Pw1a}).  It is clear from
(\ref{J}) that $J\le0$ when $c_2=\pm 1$, so the effective limits
of integration for the $c_2$ integral are given by the roots of
$J=0$.  If there are no real roots, the contribution is zero.
Specifically, $J$ has the form
 \beq J=\fr{1}{M_3^2}(A+2Bc_2-Kc_2^2), \eeq
where
 \bea &&A=(1-c_1^2)M_3^2-H^2,\nonumber\\
 &&B=(M_2+c_1M_3)H,\label{ABK}\\
 &&K=M_2^2+M_3^2+2c_1M_2M_3,\nonumber
 \eea
with
 \beq H = M_2+M_3-(1-c_1)Y, \eeq
where
 \beq Y=M_1\left(\fr{\mu}{\mu_1w_1}-1\right).\label{Y} \eeq
Note that by the triangle inequality, $Y$ is positive.  The
discriminant, which determines whether real roots exist, is
 \beq B^2+AK=(1-c_1^2)(K-H^2)=(1-c_1^2)D, \eeq
say.  If the roots are $c_{2\pm}$, then the $c_2$ integral reduces to
 \beq \int_{c_{2-}}^{c_{2+}}\fr{dc_2}{\sqrt{J}}=
 \fr{\pi M_3}{\sqrt{K}}. \eeq

Thus
 \beq P(1-w_1)=\fr{\mu M_1}{4 \mu_1 w_1^2}
 \int_{-1}^{1}dc_1\frac{(1-c_1)}{\sqrt{K}}\tha(D) \eeq

Now from (\ref{ABK})--(\ref{Y}),
 \bea D&=&(M_2+M_3)^2-2(1-c_1)M_2M_3\nonumber\\
 &&-[M_2+M_3-(1-c_1)Y]^2\nonumber\\
 &=&(1-c_1)[-2M_2M_3\nonumber\\
 &&+2(M_2+M_3)Y-(1-c_1)Y^2].\label{DY} \eea
It is useful to change variable from $c_1$ to $\xi$ defined by
 \beq \xi^2=K=(M_2+M_3)^2-2(1-c_1)M_2M_3. \eeq
The range of values of $\xi$ corresponding to $-1<c_1<1$ is then
 \beq |M_2-M_3|<\xi<M_2+M_3. \eeq
For convenience, we shall assume in what follows that $\mu_2\ge
\mu_3$, so the lower limit is $M_2-M_3$.  The effect of the
condition that the square bracket in (\ref{DY}) be positive is to
require that
 \beq \xi>\xi_c=\fr{|(M_2+M_3)Y-2M_2M_3|}{Y}. \eeq
Hence
 \beq
 P(1-w_1)=\fr{\mu M_1}{4\mu_1w_1^2}\int_{\xi_-}^{\xi_+}
 \fr{\xi d\xi}{M_2M_3}\fr{(M_2+M_3)^2-\xi^2}{2M_2M_3\xi}, \label{Pw1b}
 \eeq
where
 \bea \xi_+&=&M_2+M_3,\nonumber\\
 \xi_-&=&\max(\xi_c,M_2-M_3).\eea
The integral (\ref{Pw1b}) can be rewritten as
 \beq P(1-w_1)=\fr{\mu M_1}{8\mu_1M_2^2M_3^2w_1^2}
 \int_{\xi_-}^{\xi_+}(\xi_+^2-\xi^2)d\xi, \eeq
and hence evaluated to yield the final result
 \beq
 P(1-w_1)=\fr{\mu M_1}{24\mu_1M_2^2M_3^2}
 \fr{(\xi_+-\xi_-)^2(2\xi_++\xi_-)}{w_1^2}. \eeq

We must distinguish three different ranges of values of $w_1$
corresponding to the following ranges of $Y$:
 \bea
 1.&&0<Y<M_3,\qquad \xi_-=\fr{2M_2M_3}{Y}-(M_2+M_3);\nonumber\\
 2.&&M_3<Y<M_2,\qquad\xi_-=M_2-M_3;\\
 3.&&M_2<Y,\qquad\xi_-=(M_2+M_3)-\fr{2M_2M_3}{Y};\nonumber
 \eea
It is straightforward to rewrite the distribution in terms of the
string tensions $\mu_j$ and the variable $\dot s_1$.  The values
of $\dot s_1$ corresponding to the three ranges are easily seen to
be those in (\ref{regime}), and the corresponding expressions for $P(\dot s_1)$ are given in (\ref{ranges}).

\end{document}

%% file: fig1.pstex_t
\begin{picture}(0,0)%
\includegraphics{fig1.pstex}%
\end{picture}%
\setlength{\unitlength}{4144sp}%
\begingroup\makeatletter\ifx\SetFigFont\undefined%
\gdef\SetFigFont#1#2#3#4#5{%
  \reset@font\fontsize{#1}{#2pt}%
  \fontfamily{#3}\fontseries{#4}\fontshape{#5}%
  \selectfont}%
\fi\endgroup%
\begin{picture}(7274,9644)(-34,-8772)
\put(3061,-3796){\makebox(0,0)[lb]{\smash{{\SetFigFont{20}{24.0}{\rmdefault}{\mddefault}{\updefault}{\color[rgb]{0,0,0}$x$ link}%
}}}}
\put(4080,-1296){\makebox(0,0)[lb]{\smash{{\SetFigFont{20}{24.0}{\familydefault}{\mddefault}{\updefault}{\color[rgb]{0,0,0}$\alpha$}%
}}}}
\put(6564,-8434){\makebox(0,0)[lb]{\smash{{\SetFigFont{20}{24.0}{\rmdefault}{\mddefault}{\updefault}{\color[rgb]{0,0,0}$-v$}%
}}}}
\put(161,-4871){\makebox(0,0)[lb]{\smash{{\SetFigFont{20}{24.0}{\rmdefault}{\mddefault}{\updefault}{\color[rgb]{0,0,0}$-v$}%
}}}}
\put(5047,-3283){\makebox(0,0)[lb]{\smash{{\SetFigFont{20}{24.0}{\rmdefault}{\mddefault}{\updefault}{\color[rgb]{0,0,0}$-v$}%
}}}}
\put(765,-3490){\makebox(0,0)[lb]{\smash{{\SetFigFont{20}{24.0}{\familydefault}{\mddefault}{\updefault}{\color[rgb]{0,0,0}$v$}%
}}}}
\put(6290,142){\makebox(0,0)[lb]{\smash{{\SetFigFont{20}{24.0}{\familydefault}{\mddefault}{\updefault}{\color[rgb]{0,0,0}$v$}%
}}}}
\put(6181,-4846){\makebox(0,0)[lb]{\smash{{\SetFigFont{20}{24.0}{\familydefault}{\mddefault}{\updefault}{\color[rgb]{0,0,0}$v$}%
}}}}
\put(2093,-6483){\makebox(0,0)[lb]{\smash{{\SetFigFont{20}{24.0}{\familydefault}{\mddefault}{\updefault}{\color[rgb]{0,0,0}$\mu_3$}%
}}}}
\put(4253,-6357){\makebox(0,0)[lb]{\smash{{\SetFigFont{20}{24.0}{\familydefault}{\mddefault}{\updefault}{\color[rgb]{0,0,0}$\theta$}%
}}}}
\put(778,448){\makebox(0,0)[lb]{\smash{{\SetFigFont{20}{24.0}{\rmdefault}{\mddefault}{\updefault}{\color[rgb]{0,0,0}$-v$}%
}}}}
\put(358,-8678){\makebox(0,0)[lb]{\smash{{\SetFigFont{20}{24.0}{\familydefault}{\mddefault}{\updefault}{\color[rgb]{0,0,0}$v$}%
}}}}
\put(2841,-6930){\makebox(0,0)[lb]{\smash{{\SetFigFont{20}{24.0}{\familydefault}{\mddefault}{\updefault}{\color[rgb]{0,0,0}$u$}%
}}}}
\put(3050,-521){\makebox(0,0)[lb]{\smash{{\SetFigFont{20}{24.0}{\rmdefault}{\mddefault}{\updefault}{\color[rgb]{0,0,0}$y$}%
}}}}
\put(3082,-5471){\makebox(0,0)[lb]{\smash{{\SetFigFont{20}{24.0}{\rmdefault}{\mddefault}{\updefault}{\color[rgb]{0,0,0}$y$}%
}}}}
\put(1408,-623){\makebox(0,0)[lb]{\smash{{\SetFigFont{20}{24.0}{\familydefault}{\mddefault}{\updefault}{\color[rgb]{0,0,0}$\mu_1$}%
}}}}
\put(1345,-2298){\makebox(0,0)[lb]{\smash{{\SetFigFont{20}{24.0}{\familydefault}{\mddefault}{\updefault}{\color[rgb]{0,0,0}$\mu_2$}%
}}}}
\put(4094,-1680){\makebox(0,0)[lb]{\smash{{\SetFigFont{20}{24.0}{\rmdefault}{\mddefault}{\updefault}{\color[rgb]{0,0,0}$x$}%
}}}}
\put(4028,-6647){\makebox(0,0)[lb]{\smash{{\SetFigFont{20}{24.0}{\rmdefault}{\mddefault}{\updefault}{\color[rgb]{0,0,0}$x$}%
}}}}
\put(4951,-7891){\makebox(0,0)[lb]{\smash{{\SetFigFont{20}{24.0}{\familydefault}{\mddefault}{\updefault}{\color[rgb]{0,0,0}$K_1$}%
}}}}
\put(4951,-4966){\makebox(0,0)[lb]{\smash{{\SetFigFont{20}{24.0}{\familydefault}{\mddefault}{\updefault}{\color[rgb]{0,0,0}$K_2$}%
}}}}
\put(1216,-6001){\makebox(0,0)[lb]{\smash{{\SetFigFont{20}{24.0}{\familydefault}{\mddefault}{\updefault}{\color[rgb]{0,0,0}$\mu_1$}%
}}}}
\put(1126,-7261){\makebox(0,0)[lb]{\smash{{\SetFigFont{20}{24.0}{\familydefault}{\mddefault}{\updefault}{\color[rgb]{0,0,0}$\mu_2$}%
}}}}
\put(5266,-6721){\makebox(0,0)[lb]{\smash{{\SetFigFont{20}{24.0}{\familydefault}{\mddefault}{\updefault}{\color[rgb]{0,0,0}$\mu_1$}%
}}}}
\put(5176,-5686){\makebox(0,0)[lb]{\smash{{\SetFigFont{20}{24.0}{\familydefault}{\mddefault}{\updefault}{\color[rgb]{0,0,0}$\mu_2$}%
}}}}
\end{picture}%

%% file: constraints-final-revise.bbl
\begin{thebibliography}{99}

 \bibitem{Burgess:2001fx}
  C.~P.~Burgess, M.~Majumdar, D.~Nolte, F.~Quevedo, G.~Rajesh and R.~J.~Zhang,
  JHEP {\bf 0107}, 047 (2001)
  [arXiv:hep-th/0105204].

  \bibitem{Dvali:2001fw}
  G.~R.~Dvali, Q.~Shafi and S.~Solganik,
  arXiv:hep-th/0105203.

  \bibitem{Jones:2002cv}
  N.~T.~Jones, H.~Stoica and S.~H.~H.~Tye,
  JHEP {\bf 0207}, 051 (2002)
  [arXiv:hep-th/0203163].

\bibitem{Kachru:2003aw}
  S.~Kachru, R.~Kallosh, A.~Linde and S.~P.~Trivedi,
  Phys.\ Rev.\ D {\bf 68}, 046005 (2003)
  [arXiv:hep-th/0301240].

\bibitem{Kachru:2003sx}
  S.~Kachru, R.~Kallosh, A.~Linde, J.~M.~Maldacena, L.~McAllister and S.~P.~Trivedi,
  JCAP {\bf 0310}, 013 (2003)
  [arXiv:hep-th/0308055].

 \bibitem{Dvali:2003zj}
  G.~Dvali and A.~Vilenkin,
  JCAP {\bf 0403}, 010 (2004)
  [arXiv:hep-th/0312007].

\bibitem{Copeland:2003bj}
E.~J.~Copeland, R.~C.~Myers and J.~Polchinski,
JHEP {\bf 0406}, 013 (2004) [arXiv:hep-th/0312067].

\bibitem{Pol}
J.~Polchinski,
AIP Conf.\ Proc.\  {\bf 743}, 331 (2005) [Int.\ J.\ Mod.\ Phys.\ A
{\bf 20}, 3413 (2005)] [arXiv:hep-th/0410082].

\bibitem{Firouzjahi:2006vp}
  H.~Firouzjahi, L.~Leblond and S.~H.~Henry Tye,
  JHEP {\bf 0605}, 047 (2006)
  [arXiv:hep-th/0603161].

\bibitem{Seljak:2006bg}
  U.~Seljak, A.~Slosar and P.~McDonald,
  arXiv:astro-ph/0604335.

\bibitem{ViA1}
  T.~Damour and A.~Vilenkin,
  Phys.\ Rev.\ D {\bf 64}, 064008 (2001)
  [arXiv:gr-qc/0104026].

\bibitem{ViA2}
  T.~Damour and A.~Vilenkin,
  Phys.\ Rev.\ D {\bf 71}, 063510 (2005)
  [arXiv:hep-th/0410222].

\bibitem{Irit}
  X.~Siemens, J.~Creighton, I.~Maor, S.~Ray Majumder, K.~Cannon and J.~Read,
  Phys.\ Rev.\ D {\bf 73}, 105001 (2006)
  [arXiv:gr-qc/0603115].

\bibitem{TomMark}
M.~B.~Hindmarsh and T.~W.~B.~Kibble,
Rept.\ Prog.\ Phys.\  {\bf 58}, 477 (1995) [arXiv:hep-ph/9411342].

\bibitem{ViSh}
A.~Vilenkin and E.~P.~S.~Shellard, \emph{Cosmic Strings and other
Topological Defects} (Cambridge University Press, Cambridge,
1994).

\bibitem{Jackson:2004zg}
M.~G.~Jackson, N.~T.~Jones and J.~Polchinski,
JHEP {\bf 0510}, 013 (2005) [arXiv:hep-th/0405229].

\bibitem{Hanany:2005bc}
A.~Hanany and K.~Hashimoto,
JHEP {\bf 0506}, 021 (2005) [arXiv:hep-th/0501031].


\bibitem{Achucarro:2006es}
  A.~Achucarro and R.~de Putter,
  arXiv:hep-th/0605084.

\bibitem{Sakellariadou:2004wq}
  M.~Sakellariadou,
  JCAP {\bf 0504}, 003 (2005)
  [arXiv:hep-th/0410234].

\bibitem{Avgoustidis:2005nv}
A.~Avgoustidis and E.~P.~S.~Shellard,
arXiv:astro-ph/0512582.


\bibitem{Tong}
K.~Hashimoto and D.~Tong,
JCAP {\bf 0509}, 004 (2005) [arXiv:hep-th/0506022].


 \bibitem{Vachaspati:1986cc}
T.~Vachaspati and A.~Vilenkin,
Phys.\ Rev.\ D {\bf 35}, 1131 (1987).

\bibitem{Spergel:1996ai}
D.~Spergel and U.~L.~Pen,
Astrophys.\ J.\  {\bf 491}, L67 (1997) [arXiv:astro-ph/9611198].

\bibitem{Martins:2004vs}
  C.~J.~A.~Martins,
  Phys.\ Rev.\ D {\bf 70}, 107302 (2004)
  [arXiv:hep-ph/0410326].

\bibitem{Tye:2005fn}
S.~H.~Henry Tye, I.~Wasserman and M.~Wyman,
Phys.\ Rev.\ D {\bf 71}, 103508 (2005); \emph{ibid.}\ {\bf 71},
129906(E) (2005) [arXiv:astro-ph/0503506].

\bibitem{Copeland:2005cy}
E.~J.~Copeland and P.~M.~Saffin,
JHEP {\bf 0511}, 023 (2005) [arXiv:hep-th/0505110].

\bibitem{Saffin:2005cs}
  P.~M.~Saffin,
  JHEP {\bf 0509}, 011 (2005)
  [arXiv:hep-th/0506138].

  \bibitem{Hindmarsh:2006qn}
  M.~Hindmarsh and P.~M.~Saffin,
  JHEP {\bf 0608}, 066 (2006)
  [arXiv:hep-th/0605014].

\bibitem{Bucher:1998mh}
  M.~Bucher and D.~N.~Spergel,
  Phys.\ Rev.\ D {\bf 60}, 043505 (1999)
  [arXiv:astro-ph/9812022].

  \bibitem{Battye:1999eq}
  R.~A.~Battye, M.~Bucher and D.~Spergel,
  arXiv:astro-ph/9908047.

\bibitem{Battye:2005hw}
  R.~A.~Battye, B.~Carter, E.~Chachoua and A.~Moss,
  Phys.\ Rev.\ D {\bf 72}, 023503 (2005)
  [arXiv:hep-th/0501244].

\bibitem{PinaAvelino:2006ia}
  P.~Pina Avelino, C.~J.~A.~Martins, J.~Menezes, R.~Menezes and J.~C.~R.~Oliveira,
  Phys.\ Rev.\ D {\bf 73}, 123519 (2006)
  [arXiv:astro-ph/0602540].

\bibitem{Avelino:2006xy}
  P.~P.~Avelino, C.~J.~A.~Martins, J.~Menezes, R.~Menezes and J.~C.~R.~Oliveira,
  Phys.\ Rev.\ D {\bf 73}, 123520 (2006)
  [arXiv:hep-ph/0604250].

\bibitem{Carter:2006cf}
  B.~Carter,
  arXiv:hep-ph/0605029.

\bibitem{us}
  E.~J.~Copeland, T.~W.~B.~Kibble and D.~A.~Steer,
  Phys.\ Rev.\ Lett.\  {\bf 97} (2006) 021602
  [arXiv:hep-th/0601153].

\bibitem{Sharov:1998vy}
  G.~S.~Sharov,
  arXiv:hep-ph/9809465.

\bibitem{'tHooft:2004he}
G.~'t Hooft,
arXiv:hep-th/0408148.

\bibitem{Fomin:2000tv}
  P.~I.~Fomin and Yu.~V.~Shtanov,
  Class.\ Quant.\ Grav.\  {\bf 19}, 3139 (2002)
  [arXiv:hep-th/0008183].

\bibitem{Artru}
  X.~Artru,
  Nucl.\ Phys.\  B {\bf 85} (1975) 442.

\bibitem{BK}
  L.~M.~A.~Bettencourt and T.~W.~B.~Kibble,
  Phys.\ Lett.\  B {\bf 332} (1994) 297
  [arXiv:hep-ph/9405221].

\bibitem{Eto:2006db}
  M.~Eto, K.~Hashimoto, G.~Marmorini, M.~Nitta, K.~Ohashi and W.~Vinci,
  arXiv:hep-th/0609214.

\bibitem{RS}
R.~J.~Rivers and D.~A.~Steer, {\it Statistical mechanics of
strings with Y junctions}, work in progress, 2006.


\end{thebibliography}
